\documentclass{natureprintstyle}
\usepackage{epsfig,caption}
\usepackage{color}
\usepackage{bm}
\usepackage{graphicx}
\usepackage{longtable}
\usepackage{amssymb}
\usepackage{rotating,xcolor}
\usepackage{hyperref}

\title{All-Sky Dynamical Response of the Galactic Halo to the Large Magellanic
  Cloud}

\author{Charlie Conroy$^1$, Rohan P. Naidu$^1$, Nicol\'as Garavito-Camargo$^2$,
  Gurtina Besla$^2$, Dennis Zaritsky$^2$, Ana Bonaca$^1$, Benjamin D. Johnson$^1$}

\begin{document}

\maketitle

\let\thefootnote\relax\footnote{

\begin{affiliations}
\item Center for Astrophysics $|$ Harvard \& Smithsonian,  Cambridge, MA, USA

\item Steward Observatory, University of Arizona, 933 North Cherry Avenue, Tucson, AZ 85721, USA
  
\end{affiliations}
}

\vspace{-3.5mm}
\begin{abstract}
  
  Gravitational interactions between the Large Magellanic Cloud (LMC)
  and the stellar and dark matter halo of the Milky Way are expected
  to give rise to disequilibrium phenomena in the outer Milky
  Way\cite{Laporte18b, Garavito-Camargo19, Petersen20,
    Erkal20,Cunningham20,Tamfal20,Garavito-Camargo20}.  A local wake
  is predicted to trail the orbit of the LMC, while a large-scale
  over-density is predicted to exist across a large area of the
  northern Galactic hemisphere.  Here we present the detection of both
  the local wake and Northern over-density (hereafter the ``collective
  response'') in an all-sky star map of the Galaxy based on 1301 stars
  at $60< R_{\rm gal}<100$ kpc.  The location of the wake is in good
  agreement with an N-body simulation that includes the dynamical
  effect of the LMC on the Milky Way halo.  The density contrast of
  the wake and collective response are both stronger in the data than
  in the simulation.  The detection of a strong local wake is
  independent evidence that the Magellanic Clouds are on their first
  orbit around the Milky Way.  The wake traces the path of the LMC,
  which will provide insight into the orbit of the LMC, which in turn
  is a sensitive probe of the mass of the LMC and the Milky Way. These
  data demonstrate that the outer halo is not in dynamical
  equilibrium, as is often assumed. The morphology and strength of the
  wake could be used to test the nature of dark matter and gravity.

\end{abstract}

\vspace{1cm}


We combined optical photometry from {\it Gaia} Early Data Release
3\cite{GaiaEDR3} and infrared photometry from {\it
  WISE}\cite{Schlafly19} to identify a pure sample of giant stars
across the entire sky, excluding the Galactic plane ($|b|<10^\circ$).
Photometric distances were estimated with a 10 Gyr [Fe/H]$=-1.5$
isochrone\cite{Choi16}.  Distances were converted to Galactic physical
coordinates, and the final sample was selected to lie at
Galactocentric distances of $60< R_{\rm gal}<100$ kpc, where
simulations predict a strong signal due to the dynamical response of
the LMC and where contamination from previously known structures is
minimized.  Selections in sky coordinates and {\it Gaia} proper
motions were used to remove stars associated with known objects
including the LMC and SMC, Milky Way disk stars, and the Sagittarius
stream (see Methods for details).

\begin{figure*}
\includegraphics[width=0.5\textwidth]{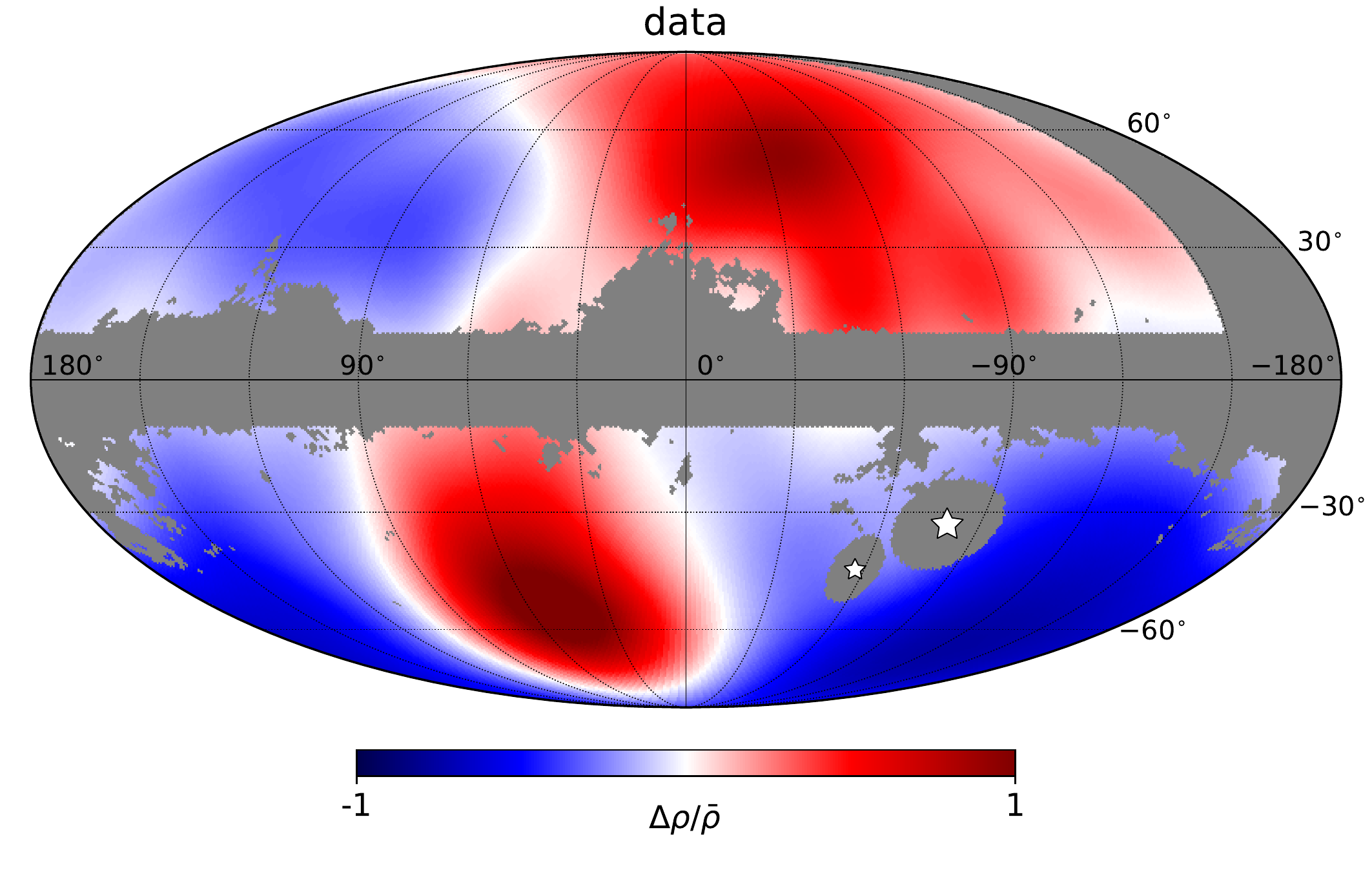}
\includegraphics[width=0.5\textwidth]{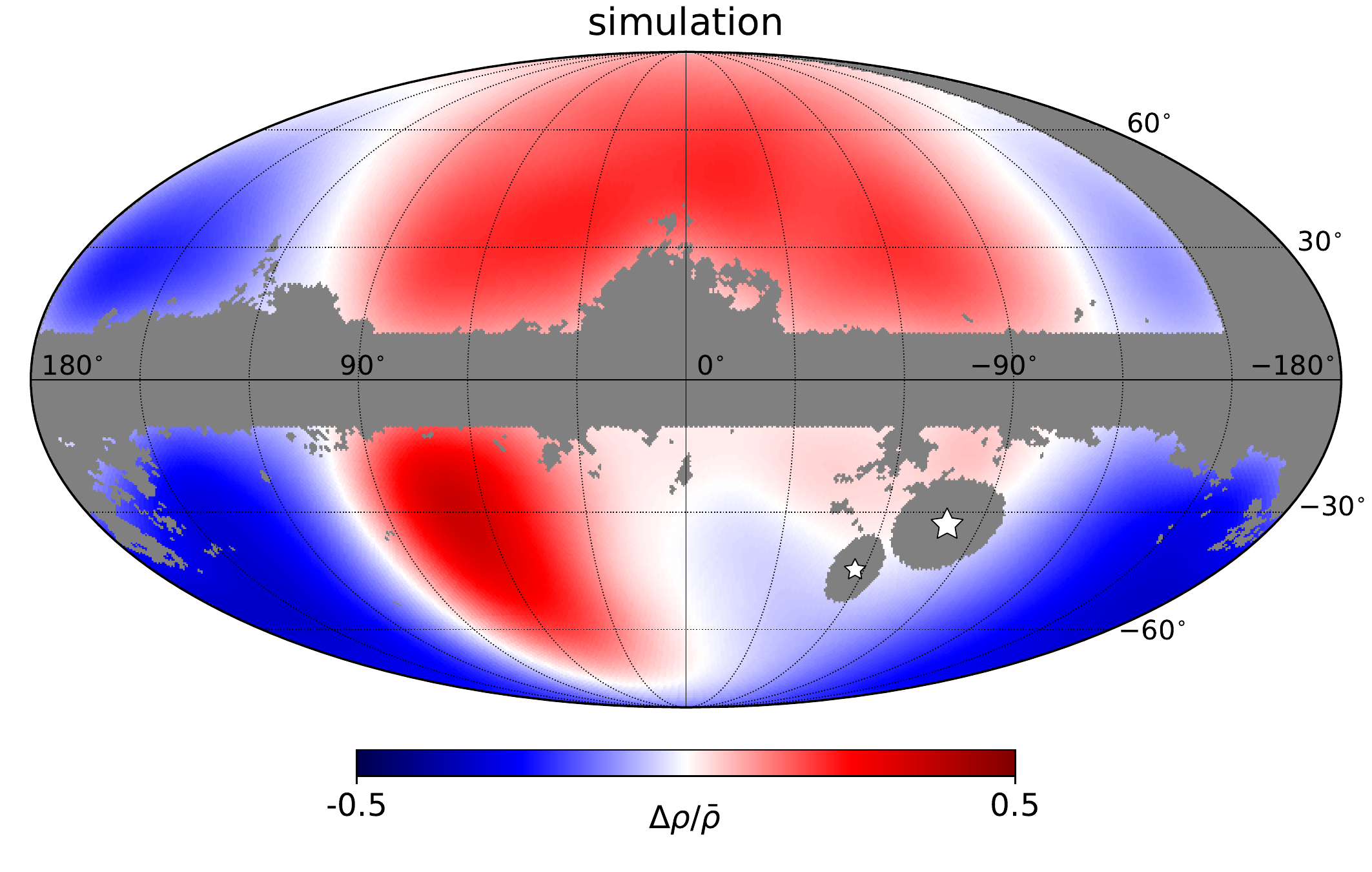}
\vspace{-4mm}
\caption{\textbf{Distribution of stars in the Galactic halo.}  All-sky
  Mollweide projection maps of the density of stars at 60 kpc
  $<R_{\rm gal}<100$ kpc, smoothed by a Gaussian kernel with
  FWHM$ =30^\circ$.  {\bf Left panel:} data based on K giant stars.
  {\bf Right panel:} simulation that includes the dynamical response
  of the halo to the orbit of the LMC.  Regions near the Galactic
  mid-plane ($|b|<10^\circ$) and the LMC and SMC, as well as regions
  of high extinction are masked in grey, while the positions of the
  LMC and SMC are marked as white stars.  One arm of the Sagittarius
  stream has been masked in sky coordinates, as indicated by the grey
  region at $b>0^\circ$ and $-180^\circ<l<-150^\circ$.  The
  overdensity in the south-west is the local wake caused by the
  passage of the LMC.  The overdensity in the North is the collective
  response of the Galaxy due to the arrival of the LMC.  Notice that
  the dynamic range of the density contrast for the simulated map is
  smaller than the observed map.}
\vspace{-4mm}
\end{figure*}

\begin{figure}[t!]
  \center
  \includegraphics[width=0.48\textwidth]{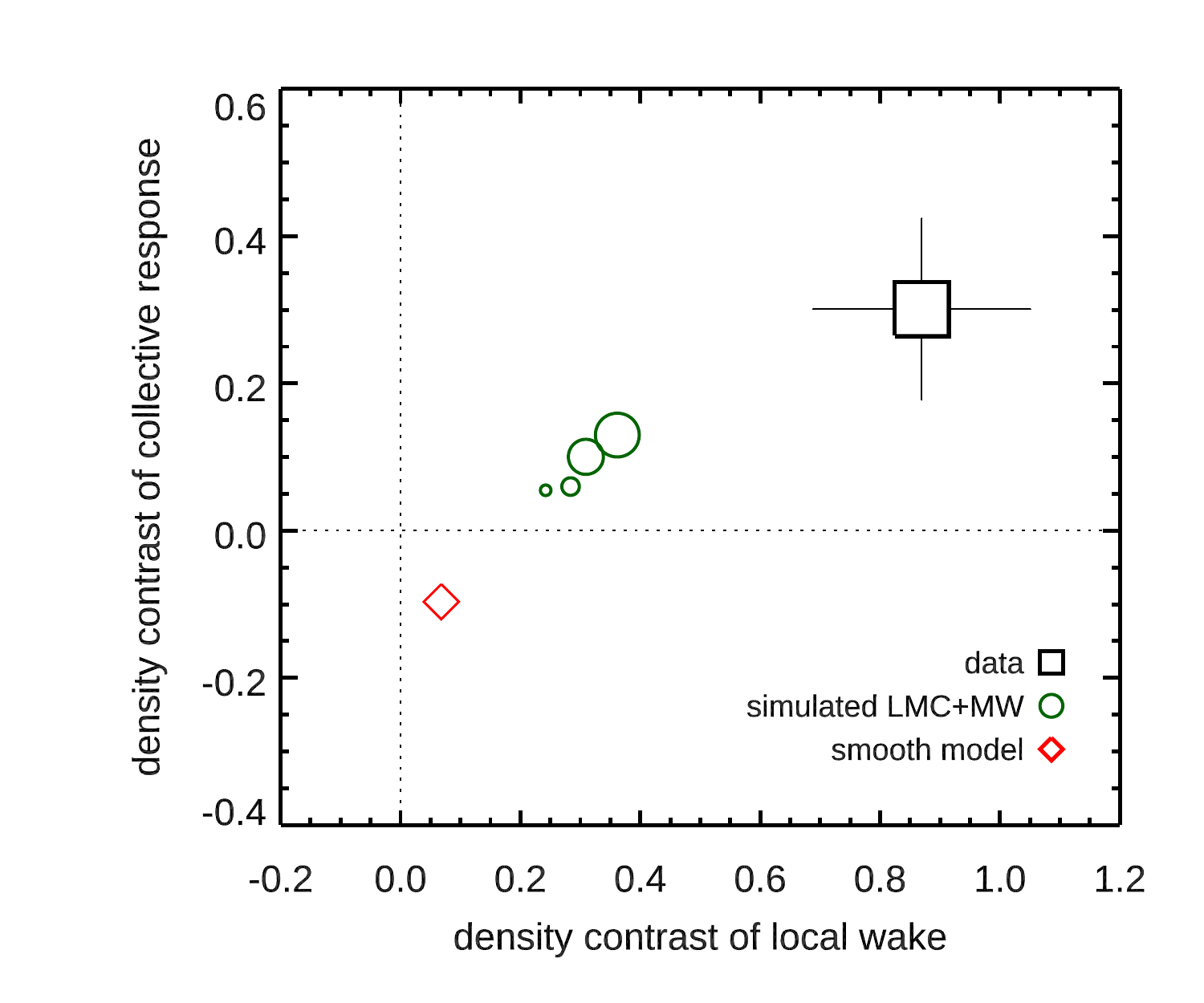}
  \caption{\textbf{Quantitative comparison between data and models.}
    Mean density contrasts in regions encompassing the local wake
    ($-60^\circ<b<-20^\circ$, $45^\circ<l<90^\circ$) and the
    collective response ($30^\circ<b<60^\circ$,
    $240^\circ<l<300^\circ$) are shown along the $x$ and $y$ axes.
    There are four simulations spanning a range in LMC masses from
    $0.8-2.5\times10^{11}M_\odot$ (small-to-large green circles).  The
    smooth model\cite{Rybizki18} is close to zero along both axes as
    expected.  The dotted lines indicate uniform distributions (no
    density contrast) on the x and y axes.  The density enhancement of
    the local wake is measured at a significance of $5\sigma$ (error
    bars on the data represent $1\sigma$ Poisson uncertainties) and
    the mean density is $1.4\pm0.2$ times larger in the data than in
    the fiducial simulation.}
\end{figure}

In Fig. 1a we show an equal-area Mollweide projection map of the
resulting sample of 1301 stars in Galactic coordinates.  The map
has been smoothed by a Gaussian beam with FWHM $=30^\circ$ and colored
by the density contrast.  Grey regions indicate portions of the sky
that have been masked.  There are two significant overdensities
spanning thousands of square degrees.  The southern feature is
strongest at $l>0^\circ$ but appears to connect directly to the LMC
and SMC (as can be seen more clearly in Extended Data Fig. 3).  The
overdensity in the north spans nearly one quarter of the entire sky.
A portion of the southern feature was previously identified as the
Pisces Plume\cite{Belokurov19a}, but the map here uncovers its full
extent on the sky.

In Fig. 1b we show predictions from an N-body simulation that includes
the dynamical response of the Galactic halo to the
LMC\cite{Garavito-Camargo19}.  The LMC orbit in this model matches
existing constraints on the present-day position and motion of the
LMC\cite{Kallivayalil13}, and has the LMC on its first passage around
the Galaxy.  The model shown here assumed a total Milky Way mass of
$M_{\rm halo}=1.2\times10^{12}\,M_{\odot}$ and an initial LMC mass of
$M_{\rm halo}=1.8\times10^{11}\,M_{\odot}$.  The simulated halo has
been processed to match the selections applied to the data, enabling a
direct comparison between the two maps.  The dynamical response of the
Galactic halo to the LMC has two primary components: 1) in the south,
a local wake is excited behind the orbit of the LMC.  This is the
classic Chandrasekhar dynamical friction wake\cite{Chandrasekhar43}.
2) A global, collective response is also created, and is mostly the
result of the movement of the barycenter of the Galaxy in the presence
of the LMC\cite{Gomez15, Garavito-Camargo19,Petersen20}.  This
collective response manifests as a large-scale overdensity in the
northern sky.  An important consequence of the simulation is that the
dynamical response should be manifest in {\it both} the stars and the
dark matter, and so a detection of structure in the former suggests a
similar level of structure in the latter.

There are several immediate implications of Fig 1.  First, the outer halo
of our Galaxy is in a state of significant disequilibrium, with order
unity variations in the density spanning thousands of square degrees.
Most previous work attempting to constrain the outer mass distribution
in the Galaxy has by necessity assumed simple equilibrium
models\cite{Williams17, Deason19a}.  Future studies must account for
the disequilibrium now measured in the Galactic halo.  Second, the very
strong observed local wake is independent evidence that the Magellanic
Clouds are on their first passage around the Galaxy.  Previous work
analyzing the positions and velocities of the Clouds had also inferred
a first-passage scenario for the orbits of the Clouds\cite{Besla07},
although those results are sensitive to the uncertain mass of the
Galaxy. If the Clouds had undergone more than one complete orbit, the
local wake would be much weaker, and perhaps not even detectable, due
to destructive interference after repeated orbits, as in the case of
the Sagittarius dwarf galaxy\cite{Garavito-Camargo19}.  Third, the
local wake is predicted to accurately trace the orbital path of the
LMC\cite{Garavito-Camargo19}, and so its observed location will
provide stringent new constraints on the orbit of the LMC.

In Fig. 2 we provide a quantitative comparison between observations
and simulations of the strength of the local wake and collective
response.  Regions encompassing the local wake
($-60^\circ<b<-20^\circ$, $45^\circ<l<90^\circ$) and the collective
response ($30^\circ<b<60^\circ$, $240^\circ<l<300^\circ$) are defined
and the mean density contrast is measured within them (using
unsmoothed versions of the maps in Fig. 1).  Poisson statistics are
used to estimate $1\sigma$ uncertainties.  In this figure we show
results for four simulations with a range of LMC masses at infall:
$(0.8,1.0,1.8,2.5)\times 10^{11}\,M_{\odot}$.  The density contrast in
the local wake increases with increasing LMC mass, but overall the
simulations predict a density contrast lower than observed.  In the
data, the collective response is largely confined to
$-180^\circ<l<0^\circ$.  Although the collective response is expected
to be broadly asymmetric\cite{Garavito-Camargo19}, in the observed
footprint the simulated collective response is approximately symmetric
about $l=0^\circ$ (see Methods for further discussion).  We also show
results for a smooth stellar halo\cite{Rybizki18} - as expected this
model shows no significant excess or deficit of stars in the wake or
collective response regions.

The simulation is a genuine prediction and was not calibrated in any
way to reproduce the observed features.  The overall agreement between
the simulation and observations, especially in the Southern
hemisphere, is therefore quite striking.  Unsurprisingly, given the
large available parameter space, the agreement is not perfect.  For
example, the density of the local wake in the fiducial
simulation is lower by a factor of $1.4\pm0.2$ and the location of the
peak density occurs further away from the LMC compared to the
observations.  In addition, the collective response in the northern
hemisphere is not as asymmetric (west vs. east) in the simulation as
in the data. The precise location and density of the local wake is
sensitive not only to the mass of the LMC, but, because the formation
of the wake is a resonant process\cite{Weinberg98}, is also sensitive
to the orbit of the LMC and the distribution of orbits within and
shape of the Milky Way halo.  Furthermore, the impact of the SMC on
these predictions has not yet been studied.  The initial simulated
stellar halo was smooth; the amplitude of the wake in a realistic halo
built from mergers might be different.  Exploration of this parameter
space will be necessary to determine if such a large observed wake
signal can be accommodated within conventional models for the Milky
Way and LMC.  Because the wake is sculpted by the total gravitational
mass and not just the stars, its existence and detailed morphology may
also provide stringent tests of non-standard dark matter (e.g., fuzzy
dark matter, self-interacting dark matter)\cite{Furlanetto02, Hui17,
  Lancaster20} and alternative gravity models\cite{Ciotti04,
  Nipoti08}.

The exact location of the collective response on the sky is also sensitive
to the orbit of the LMC, which is in turn sensitive to the mass of the
Milky Way.  As with the local wake, the collective response is
sensitive to the distribution of orbits within the Galactic
halo. Simulations that explore these variables will be necessary to
see if the observed location of the collective response can be
reproduced.

The Magellanic stream is a vast structure of cold gas trailing and
originating from the LMC and SMC\cite{DOnghia16, Lucchini20}.
Simulations predict a corresponding stellar stream in the vicinity of
the gaseous stream\cite{Gardiner96, Diaz12}, although its location and
mass content is uncertain.  The stellar stream is predicted to be
narrower on the sky than the local wake\cite{Garavito-Camargo19} and
to reside at $R_{\rm gal}>100$ kpc at the location of the local
wake\cite{Besla13}.  There have recently been detections of cold
stellar structure in this region of the sky, perhaps associated with
the Magellanic system in some way\cite{Deason18b, Zaritsky20}.  In the
Methods we provide an un-smoothed version of Fig. 1a which does not
reveal a colder feature that could be associated with a stream.
Kinematic information should be definitive -- a stellar stream, if one
exists, should have distinct kinematic behavior compared to the local
wake.

Simulations that include the effect of the LMC on the Galactic
halo\cite{Garavito-Camargo19, Cunningham20, Petersen20} predict large
radial and tangential velocity differences across the sky due to the
dynamical response of the halo to the LMC, with amplitudes exceeding
$45\, {\rm km\,s}^{-1}$.  The sample presented here will be ideal for
measuring the velocity signal from spectroscopy and proper motions, as
nearly all stars are brighter than $G=17.5$.  Independent work has
recently detected the predicted velocity signatures\cite{Petersen20b,
  Erkal20b}.  These detections provide strong corroboration that the
density variations reported here are due to the dynamical response of
the Galaxy to the presence of the Magellanic Clouds.  The joint
mapping and modeling of the density and kinematic signatures is the
next step in uncovering the complex phase space structure imprinted by
the Magellanic Clouds in the outer Galaxy.


\begin{addendum}
  
\item [Acknowledgements] We thank both referees for their constructive
  feedback.  C.C. is partially supported by the Packard
  Foundation.  R.P.N. gratefully acknowledges an Ashford Fellowship
  and Peirce Fellowship granted by Harvard University.  G.B.  and
  N.G.-C. are supported by HST grant AR 15004, NASA ATP grant
  17-ATP17-0006, NSF CAREER AST-1941096.  A.B. acknowledges support
  from NASA through HST grant HST-GO-15930.  All the simulations were
  run on El-Gato super computer which was supported by the National
  Science Foundation under Grant No. 1228509.  We have made use of
  data from the European Space Agency mission Gaia
  (http://www.cosmos.esa.int/gaia), processed by the Gaia Data
  Processing and Analysis Consortium (DPAC; see
  http://www.cosmos.esa.int/web/gaia/dpac/consortium). Funding for
  DPAC has been provided by national institutions, in particular the
  institutions participating in the Gaia Multilateral Agreement.  This
  publication makes use of data products from the {\it Wide-field
    Infrared Survey Explorer}, which is a joint project of the
  University of California, Los Angeles, and the Jet Propulsion
  Laboratory/California Institute of Technology, and NEOWISE, which is
  a project of the Jet Propulsion Laboratory/ California Institute of
  Technology.  {\it WISE} and NEOWISE are funded by the National
  Aeronautics and Space Administration.

\item[Author Contributions] C.C. and R.P.N. jointly conceived of the
  project.  C.C. led the analysis of the data.  N.G-C. and
  G.B. provided the simulation data and aided in its interpretation.
  All authors contributed to aspects of the analysis and to the
  writing of the manuscript.

\item[Author  Information]  Reprints and permissions
    information is available at npg.nature.com/reprintsandpermissions.
    Correspondence and requests for materials should be addressed to
    C.C.\ (cconroy@cfa.harvard.edu).

  \item[Data Availability] The K giant catalog used in this paper is
    available at \texttt{https://doi.org/10.7910/DVN/2D1H8J}.
    
  \item[Code Availability] We have opted not to make the code used in
    this manuscript available because the data reduction and analysis
    is straightforward and can be easily reproduced following the
    methods described herein.
    
\end{addendum}


\newpage

\setcounter{page}{1}
\setcounter{figure}{0}
\setcounter{table}{0}
\captionsetup[figure]{labelformat=empty}
\renewcommand{\thefigure}{Extended Data \arabic{figure}}
\renewcommand{\thetable}{Extended Data \arabic{table}}

\begin{center}
{\bf \Large \uppercase{Methods} }
\end{center}

\noindent
{\bf Identification of giants}
\\
\noindent
In order to study the outer stellar halo across the entire sky we need
a means of identifying, with high purity, a sample of giant stars
based solely on all-sky photometry.  Previous work\cite{Majewski03,
  Koposov15} employed 2MASS $JHK_s$ photometry to identify M giant
stars by relying on the pressure-sensitivity of continuous opacity
sources (mostly $H^-$).  However, the relatively shallow depth of
2MASS precluded a detailed view of the outer halo (our sample of stars
is largely confined to $K_s>13$ while previous work was limited to
$K_s<13$).  We recently presented a photometric selection technique
\cite{Conroy18a} based on Pan-STARRS and {\it WISE} colors that
efficiently identified K giants to W1 $=15$.  The downside to this
approach is that Pan-STARRS is not an all-sky survey.  We therefore
decided to explore the possibility of using {\it Gaia} and {\it WISE}
photometry in order to select giants across the entire sky.  In order
to correct for dust extinction we adopt the following reddening
coefficients: $A_{BP}/E(B-V)=3.0$, $A_{RP}/E(B-V)=1.92$, and
$A_{W1}/E(B-V)=0.18$.  The normalization of the standard $E(B-V)$ dust
map\cite{Schlegel98} is reduced by 14\% following more recent
work\cite{Schlafly11}. We also make two selections on {\it Gaia}
quality flags: \texttt{ruwe}$<1.4$ and $3\sigma$ clipping of the
corrected BP and RP flux excess factor $C^*$\cite{Riello20}.  We also
apply a parallax selection of $\pi<0.2$ mas to remove obvious
foreground stars, using the corrected parallaxes from {\it
  Gaia}\cite{Lindegren20b}.  For the WISE data, we restrict to
$12<W1<15$, a magnitude range that is nearly 100\%
complete\cite{Schlafly19} and has formal uncertainties of 0.01 mag at
the faint end of this range.

Extended Data Fig. 1 shows the distribution in $BP-RP$ and $RP-W1$ colors
for stars with $12<W1<15$ at high Galactic latitudes.  This specific
subsample was chosen to minimize Galactic reddening.  Two sequences
are clearly visible, with the upper and lower branches containing
giants and dwarfs, respectively.  We fit a polynomial to the giant
sequence:
\begin{equation}
(RP-W1)_{\rm fid} = -0.9134+2.5985\, (BP-RP)-0.4518\,(BP-RP)^2
\end{equation}
and the red lines are defined by $(RP-W1)_{\rm fid}+0.06$ and $(RP-W1)_{\rm
  fid}-0.05$.  We also impose a limit of $1.8<(RP-W1)<2.5$.  This
selection is applied to an all-sky catalog of {\it Gaia} and {\it
  WISE} cross-matched sources.    For a metallicity of [Fe/H]$=-1.5$,
this color range corresponds to red giants with $3800<T_{\rm eff}<4400$ K,
i.e., K giant stars.

\begin{figure}[t!]
  \includegraphics[width=0.45\textwidth]{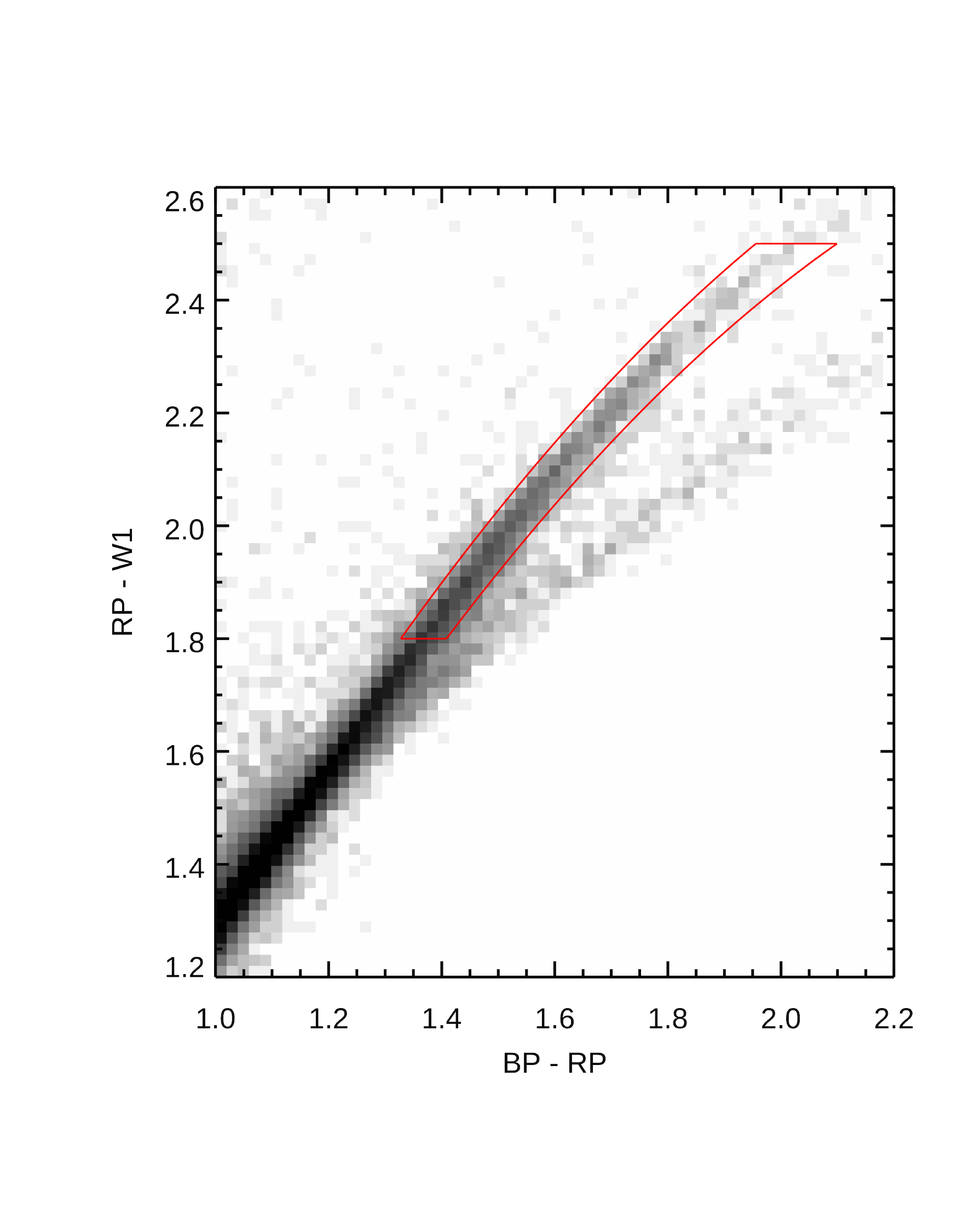}
  \vspace{-1cm}
  \caption{\textbf{Extended Data Figure 1 $|$ Photometric selection of
      giants.}  Color-color diagram for stars at high Galactic
    latitude ($b>45^\circ$) with {\it Gaia} parallax $\pi<0.1$ mas.
    Two sequences are clearly visible, with the upper branch being
    associated with giant stars, and the lower with dwarf stars. The
    red lines mark the selection boundary used in this work.}
\end{figure}

\begin{figure}[t!]
  \includegraphics[width=0.45\textwidth]{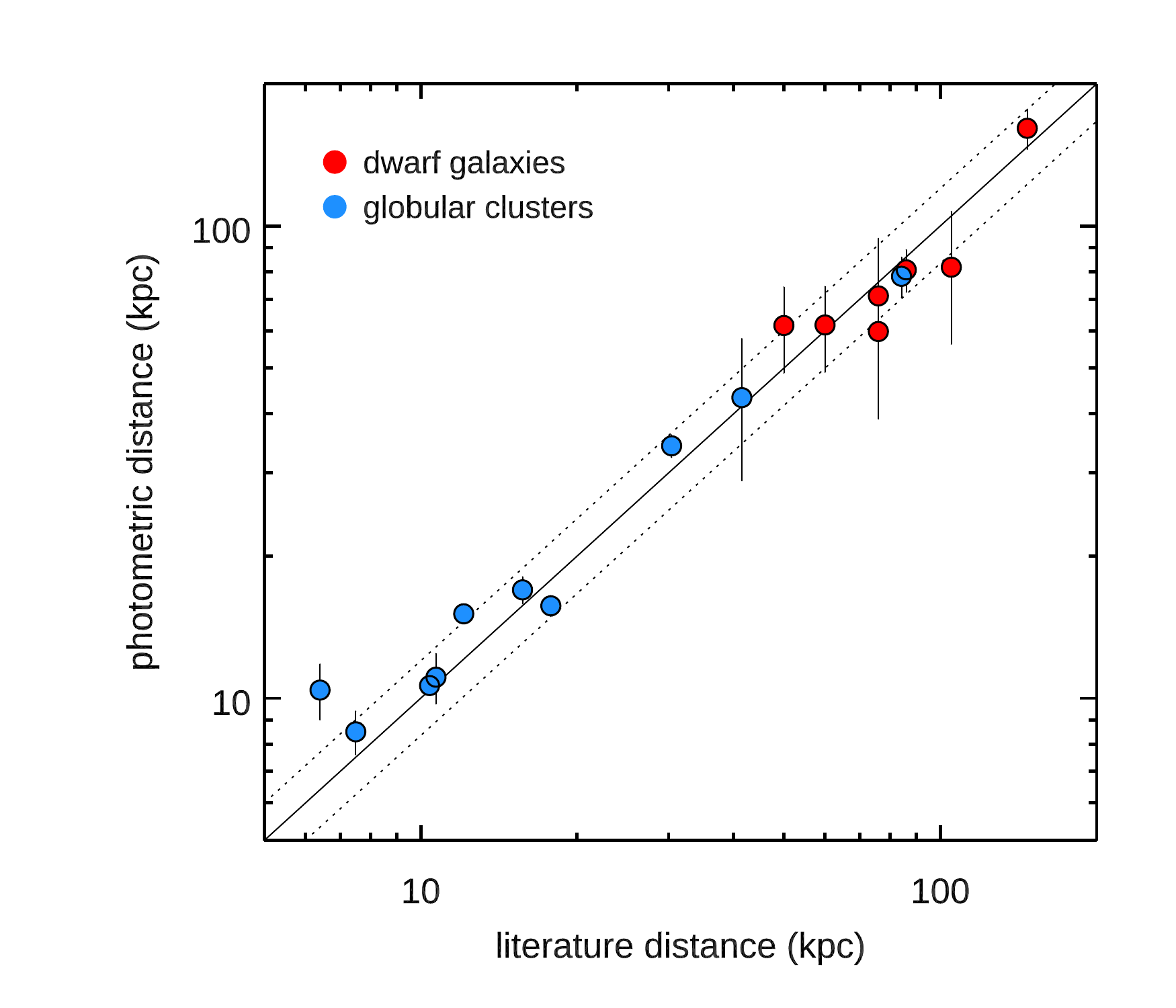}
  \caption{\textbf{Extended Data Figure 2 $|$ Test of Photometric
      Distances.}  Comparison of our photometric distances against
    literature values for satellite galaxies around the Milky Way
    (red) and globular clusters (blue).  Error bars represent the
    $1\sigma$ scatter in the photometric distances and dotted lines
    mark $\pm20$\% about the one-to-one line.  The dwarf galaxies
    include Fornax, Draco, Sculptor, Carina, Ursa Minor, and the LMC
    and SMC, and span average metallicities from [Fe/H]$\approx-0.5$
    (LMC) to $-2.2$ (Ursa Minor)\cite{McConnachie12}.}
\end{figure}

Distances are estimated for these giants by using \texttt{MIST}
stellar isochrones\cite{Choi16}.  Specifically, we select red giant
branch stars from a 10 Gyr, [Fe/H]$=-1.5$ model and fit a quadratic
function between the $BP-RP$ color and the $W1$ absolute magnitude:
\begin{equation}
M_{W1} = 11.547 -17.117\, (BP-RP) + 3.9329\, (BP-RP)^2
\end{equation}
This equation is used to estimate distances for all stars in the
catalog.  Distances based on a single color will not be very accurate
as good photometric distances require some knowledge of the age and
metallicity, both of which we have fixed here.  The stellar halo is
widely believed to be old, and even factors of two change in the
adopted age changes the inferred distance by only $\approx7$\%.  The
metallicity has a larger impact on the inferred distances.  For
example, assuming a fixed color of $BP-RP=1.4$, increasing the
metallicity to [Fe/H]$=-1.0$ would result in 25\% closer distances,
while assuming [Fe/H]$=-2.0$ would place the distances 14\% further
away compared to our fiducial isochrone.  We have adopted
[Fe/H]$=-1.5$ based on the fact that the mean metallicity of the halo
is $\langle$[Fe/H]$\rangle=-1.2$ with some evidence for a slightly
more metal-poor halo at $R_{\rm gal}>50$ kpc\cite{Conroy19b}.  We
emphasize that precise distances are not required in this work -- they
are simply used to place stars in broad radial bins.

The strongest test of these photometric distances comes from direct
comparison to literature distances.  This test is provided in Extended
Data Fig. 2.  We identified stars in our giant catalog that are
spatially associated with dwarf satellite galaxies and globular
clusters (GCs).  In the former category, we found significant numbers
of stars associated with Fornax, Draco, Sculptor, Carina, Ursa Minor,
the LMC and SMC.  For the GCs, we found nine objects with 10 or more
stars in our catalog with $12<W1<14$.  The figure compares the median
photometric distance for each source to known literature
values\cite{Harris96, McConnachie12}.  The agreement is good at the
$\pm20$\% level, which gives confidence that the photometric distances
are both accurate and reasonably precise.

\begin{figure*}[t!]
  \includegraphics[width=0.5\textwidth]{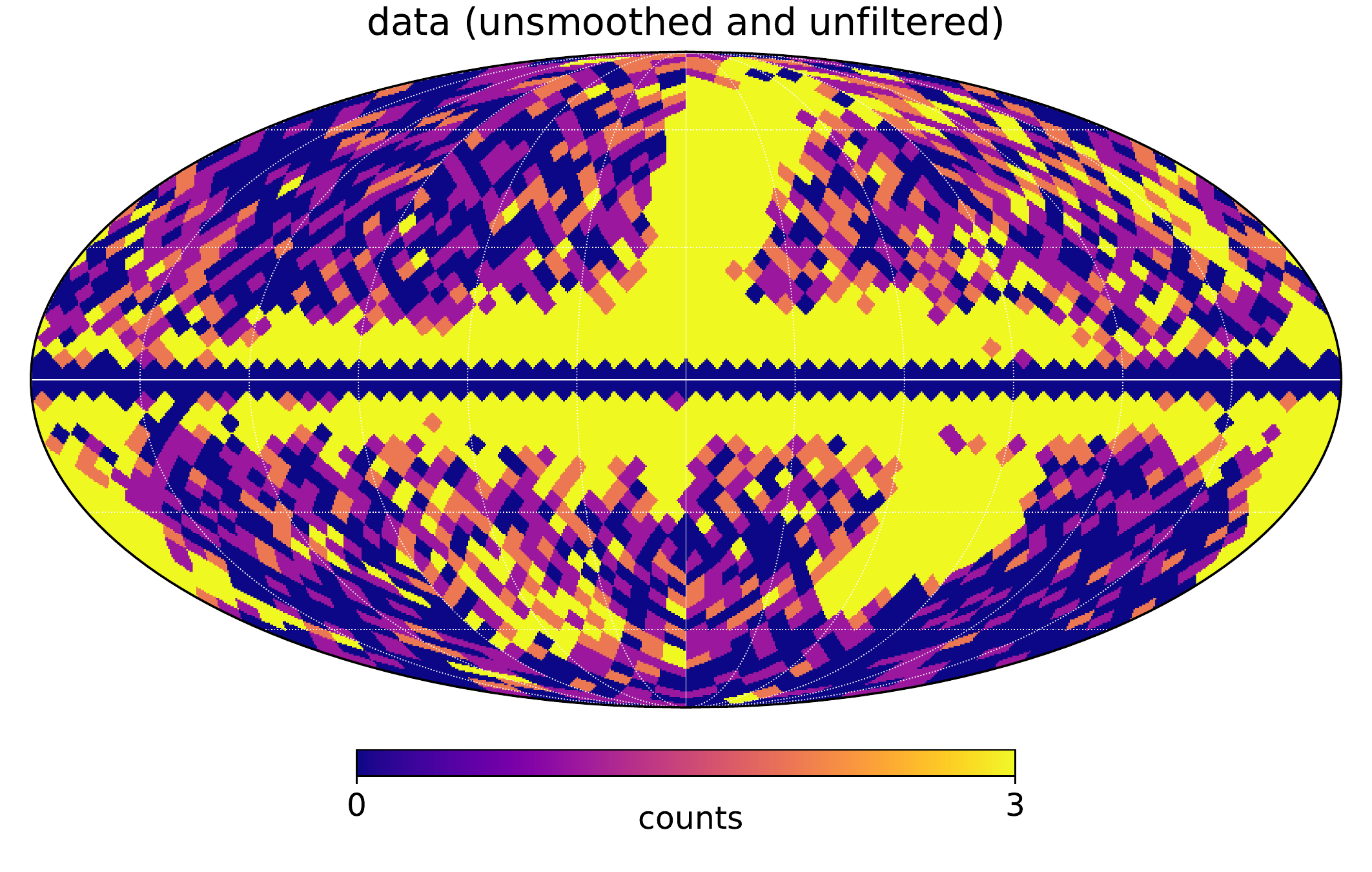}
  \includegraphics[width=0.5\textwidth]{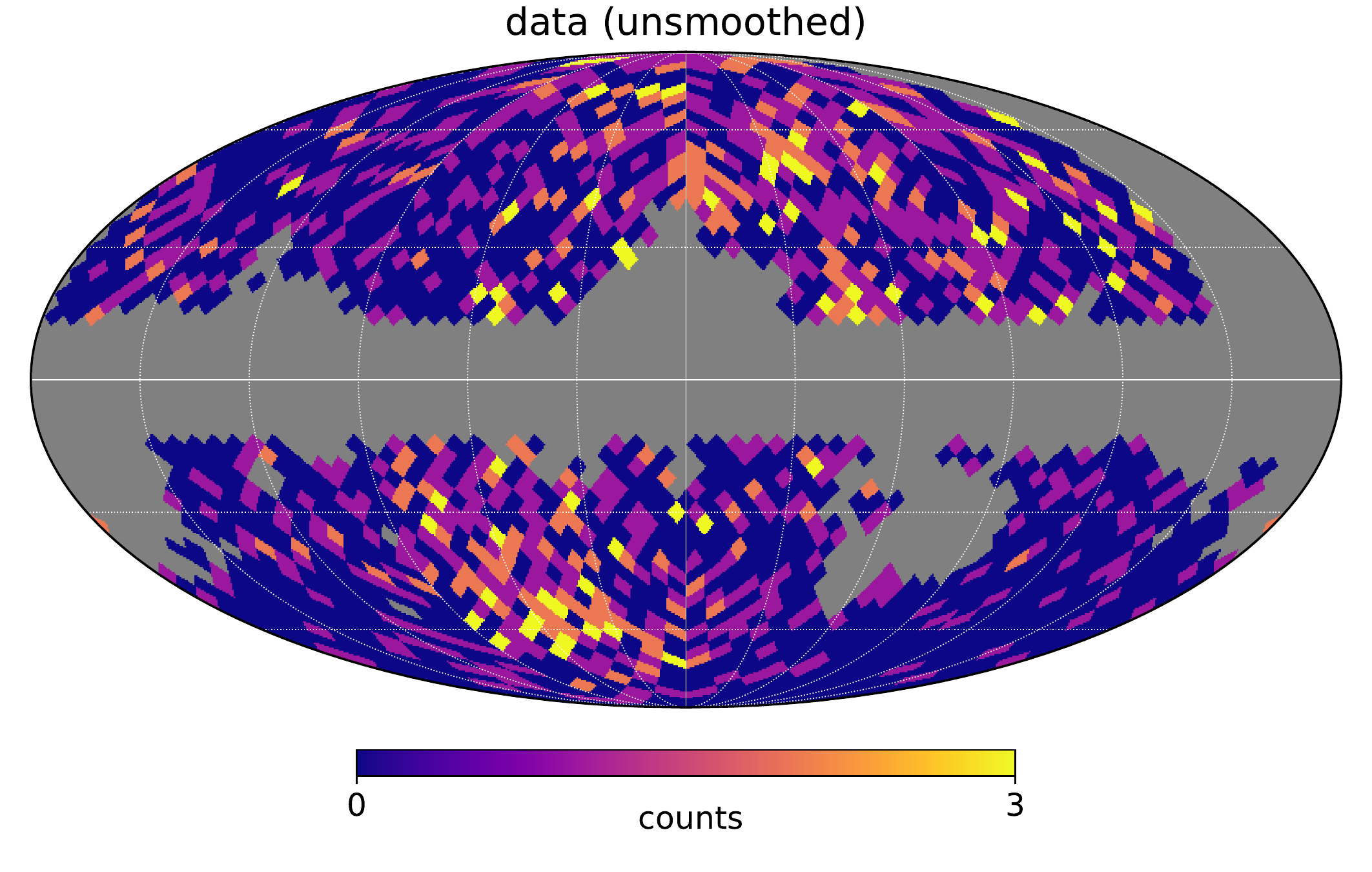}
  \caption{\textbf{Extended Data Figure 3 $|$ Maps of un-smoothed star
      counts.}  Mollweide projection maps of the observed sample of
    giants with pixels of size 13.4 sq. deg. {\bf Left panel:} map
    with no masking of known structure, either in sky coordinates or
    proper motions.  The two main features identified here, the
    transient wake and the collective response, are still clearly
    visible even in this unfiltered map.  The LMC and SMC appear in
    the lower right as a merged region of high density.  Other
    features include the stellar disk+bulge in the center, and the
    Sagittarius stream both in the north-center and lower left and
    right.  {\bf Right panel:} map showing the filtered catalog used
    in Fig 1a.}
\end{figure*}

\begin{figure*}[t!]
  \center
  \includegraphics[width=0.8\textwidth]{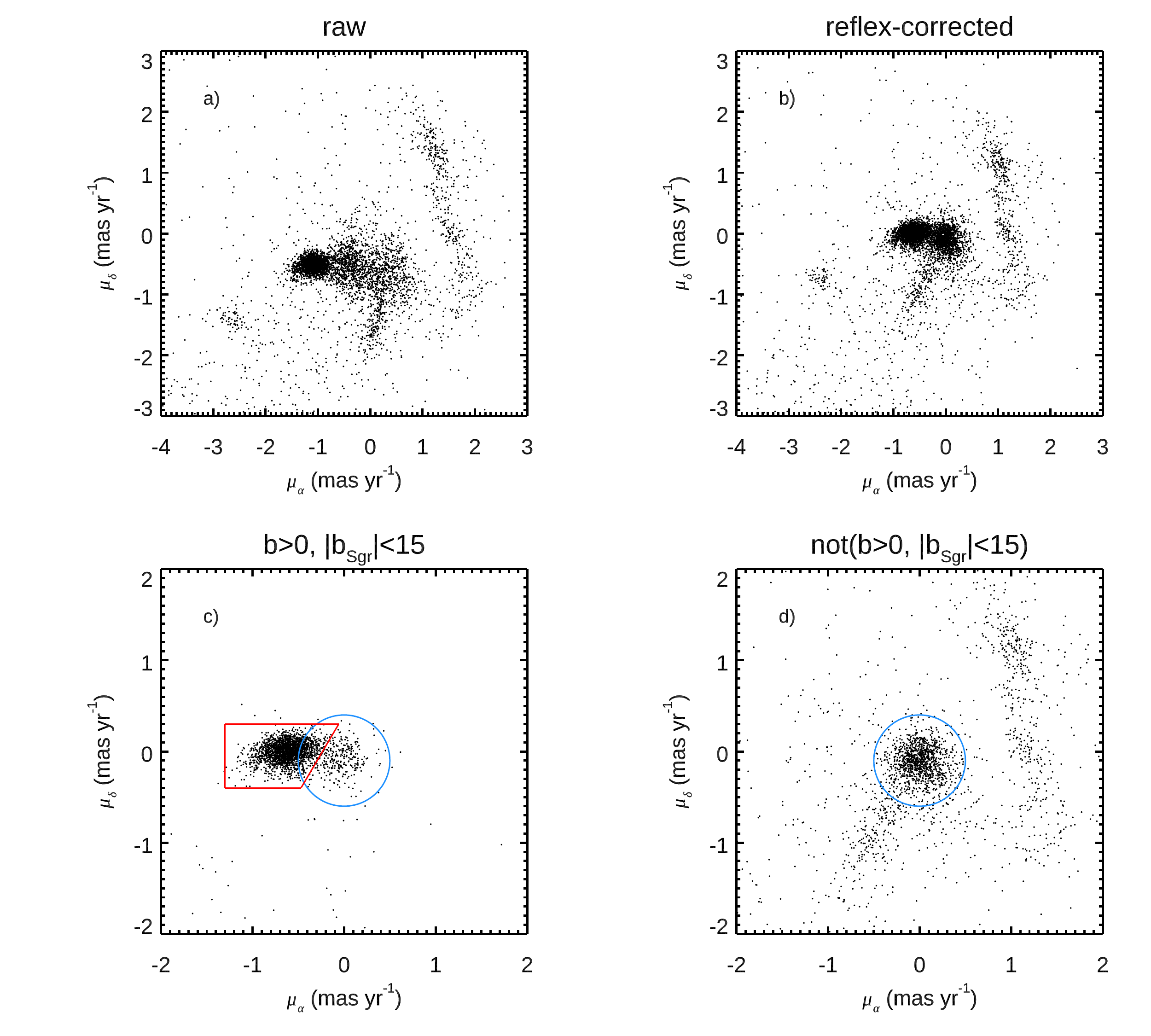}
  \caption{ \textbf{Extended Data Figure 4 $|$ Proper motions of the
      halo sample.}  {\bf a:} proper motions of the K giants at
    $60<R_{\rm gal}<100$ kpc with $12<W1<15$ and the LMC and SMC
    removed via on-sky selection.  {\bf b:} solar reflex-corrected
    proper motions.  Notice the much tighter distribution of stars
    near (0,0).  In this panel the Sagittarius dSph is visible at
    (-2.5,-0.7), LMC and SMC stars not removed by the on-sky selection
    are visible as a narrow vertical vertical strip at
    $\mu_\alpha\gtrsim0.5$ mas yr$^{-1}$.  The Northern Sagittarius
    arm is the large overdensity at (-0.7,0.0), and a Southern arm of
    Sagittarius is the diagonal cluster of points at $\mu_\delta<-0.5$
    mas yr$^{-1}$.  {\bf c:} reflex-corrected proper motions focusing
    on the region of the sky containing the Northern Sagittarius arm
    at $l\sim0^\circ$.  The red box indicates our selection for
    removing this feature.  {\bf d:} stars not in the selection shown
    in panel {\bf c}. Our selection for low proper motion stars is
    indicated by a blue circle in {\bf c, d}.}
\end{figure*}

\vspace{.3cm}
\noindent
{\bf Removal of structure and map-making}
\\
\noindent
Starting with the catalog of giants, we identify the parent sample as
stars that satisfy a magnitude selection of $12<W1<15$ and a {\it
  Gaia} parallax selection of $\pi<0.2$ mas (we use the corrected
parallaxes available in {\it Gaia} EDR3\cite{Lindegren20b}).  We also
remove stars with $E(B-V)>0.3$ (regions with such large reddening are
masked in the all sky maps).  We then select stars with Galactocentric
distances of $60<R_{\rm gal}<100$ kpc, and further require stars to
lie off the Galactic plane ($|b|>10^\circ$).  We also remove stars
clearly associated with the dwarf galaxies and GCs mentioned above.
This sample comprises 146,926 stars (the vast majority of which are
associated with the LMC and SMC).  Extended Data Fig. 3a shows the
density distribution in Galactic coordinates of this sample in a
Mollweide projection.  Each pixel has an area of 13.4 sq. deg and the
color is proportional to the number of stars in each pixel. 

There is significant structure throughout the sky.  The brightest
features in this map are associated with the LMC and SMC (lower
right), the Galactic disk and bulge (center) and the Sagittarius
stream (center-north, and at the edges of the map in the South).  The
overdensities associated with the local wake and collective response
are clearly visible in this raw map.  However, in an attempt to
isolate the features of interest we have selected various populations
for removal on the basis of sky coordinates and proper motions.
  
The LMC and SMC are removed via selections in Galactic $(l,b)$ space.
In the case of the LMC, a circular region with radius of $8^\circ$
centered on the LMC was excised from the catalog.  For the SMC, an
ellipse with semi-major and minor axes of $3.2^\circ$ and $2.5^\circ$
centered on the SMC was used to remove stars (additional LMC and SMC
stars beyond the Galactic coordinate cut are removed via the proper
motion selection below).  These selections reduce the catalog to 5,007
stars.

Proper motions offer an efficient means by which we can remove
structure unassociated with the diffuse halo.  We work with solar
reflex motion-corrected proper motions\footnote{We adopt the
  Galactocentric frame implemented in \texttt{Astropy v4.0}
  \cite{astropy} which has the following parameters: $R_{0}=8.122$ kpc
  \cite{Gravity19},
  $[V_{\rm{R,\odot}}, V_{\rm{\phi,\odot}}, V_{\rm{Z,\odot}}]=[-12.9,
  245.6, 7.78]$ km s$^{-1}$ \cite{Drimmel18}, $Z_{\rm{\odot}}=20.8$ pc
  \cite{BennettBovy19}. We use the \texttt{gala}\cite{gala} package
  \texttt{reflex\_correct} by setting radial velocities to zero to
  account for the imprint of the solar motion on our proper motions.}.
The effect of the reflect motion correction is shown in Extended Data
Fig. 4.  The upper left panel shows the raw proper motions while the
upper right shows the reflex-corrected proper motions.  The lower left
panel shows the sample at $b>0^\circ$ and $|b_{\rm Sgr}|<15^\circ$
where $b_{\rm Sgr}$ is the latitude in the frame of the Sagittarius
orbital plane\cite{Belokurov14}.  The dense clump contains the
Sagittarius stream in the North at $l\sim0^\circ$.  The red box is
used to remove these stars and is defined by $\mu_\alpha>-1.3$ mas
yr$^{-1}$, $-0.4<\mu_\delta<0.3$ mas yr$^{-1}$, and
$\mu_\delta > 1.7 \mu_\alpha+0.4$ mas yr$^{-1}$ for $b>0^\circ$ and
$|b_{\rm Sgr}|<15^\circ$.  The remaining Northern arm of the
Sagittarius stream is removed by simply masking the region $b>0^\circ$
and $180^\circ<l<210^\circ$ (it has low proper motion and so cannot be
easily isolated from the rest of the halo via proper motion selections).

The Sagittarius selections above reduce the sample to 2,744 stars.  A
final selection in proper motion space is indicated by the blue
circles in the bottom panels of Extended Data Fig. 4:
$\mu_\alpha^2 + (\mu_\delta+0.1)^2<0.5^2$ (in units of mas yr$^{-1}$).
This selection removes disk stars, LMC and SMC stars beyond the on-sky
selection, the Sgr dSph, and other Sgr arms, and reduces the sample to
1301 stars.  Of the stars removed with this selection, 520 are
associated with the LMC and SMC and 78 with the Sgr dSph.  The rest
are associated with either a southern arm of Sgr or are confined to
$|b|<20^\circ$ (these are disk stars with incorrect photometric
distances, likely due to their very different metallicities).  We note
that the median proper motion uncertainty of the sample of 2,744
stars is 0.07 and only 110 stars have an uncertainty in either RA or
DEC of $>0.15$ mas yr$^{-1}$.

The final catalog of 1301 stars is available for download at
\texttt{https://doi.org/10.7910/DVN/2D1H8J}.  Extended Data Fig. 3b
shows the distribution of the final clean catalog.  Comparison between
the two panels reveals that the various selections have a small impact
on the regions of the sky containing the local wake and the collective
response.  Importantly, both of these features are visible in the raw,
un-smoothed map.  Extended Data Fig. 5 shows an annotated map of the
final sample, including the orbit of the LMC (white) and the regions
used to define the number counts used in Fig. 2.

In an effort to provide a direct comparison between the data and
models we have applied the exact same selection criteria outlined
above to the models.  In particular, the sky-based selections
(including the LMC, SMC, Galactic plane, and locations where
$E(B-V)>0.3$ are applied as a spatial mask in the maps.  Selections in
(solar reflex-corrected) proper motions are applied within the
catalogs.  An exception to this is the selection of the Northern
Sagittarius arm in proper motion space; this selection is only applied
in the data.  We do this because, while the selection is very clean in
data space, the model halo particles are somewhat kinematically
``hotter'' than the data and so this selection would remove a larger
number of model halo particles.  We also re-sample the models to
follow the same distribution in $R_{\rm gal}$ as the data, and have
applied a 20\% uncertainty to heliocentric distances to mimic our
photometric distance uncertainties.

The following steps were taken to compute smoothed maps (as in e.g.,
Fig. 1).  First, we bin the stars into pixels.  Second, we apply the
spatial mask.  Third, for the non-masked pixels we compute the average
density and divide the map by that average (and then subtract one to
compute the density contrast).  Fourth, we smooth the map by a
Gaussian kernel.  Smoothing a masked map creates challenges near the
masked regions.  As we are using a continuous kernel, we opted to
simply set the masked regions to 0.0 for the masking procedure (which
effectively means that we assume mean density for the masked regions).
This is the default behavior of the \texttt{healpy}\cite{healpy}
routines used in this work.  Finally, the mask is re-applied to the
maps after smoothing to avoid the appearance of the data ``bleeding''
into the masked regions.  We emphasize that the exact same procedure
was applied to both the data and models, enabling a robust comparison.
  
Extended Data Fig. 6 shows the cumulative effect of these selections
on a smooth model of the Milky Way\cite{Rybizki18}.  This model
includes an oblate power-law stellar halo with ellipticity
$\epsilon=0.76$ that is aligned with the disk plane.  The map in the
left panel shows the total sample, which is uniform and symmetric
about both N-S and E-W axes.  The concentration of stars toward the
mid-plane is a reflection of the oblate shape of the stellar halo.
The map in the right panel shows the model after the selections have
been applied.  The proper motion selection has a minor effect on the
spatial distribution of the smooth model.

\begin{figure}[t!]
  \center
  \includegraphics[width=0.5\textwidth]{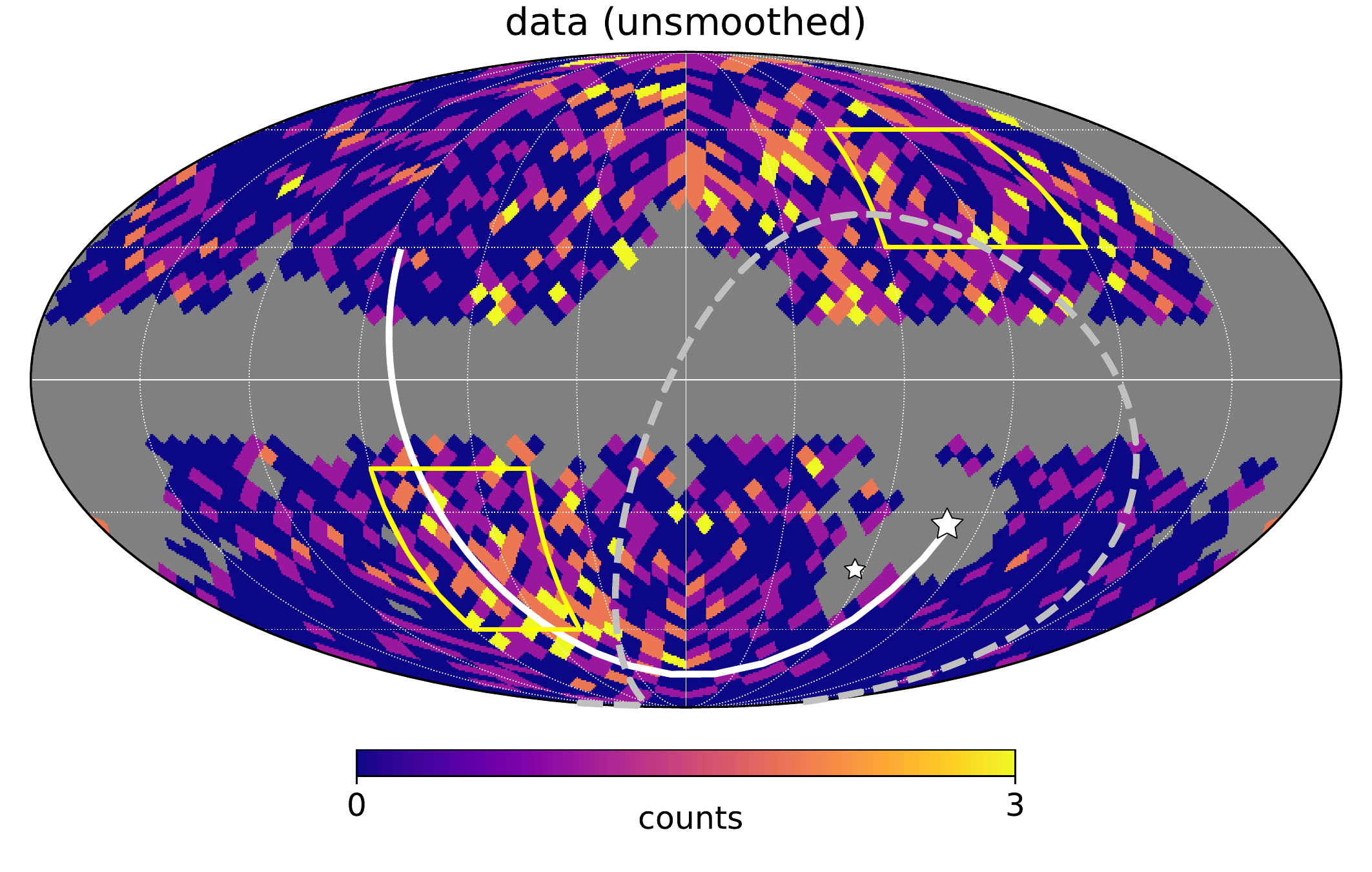}
  \caption{ \textbf{Extended Data Figure 5 $|$ An annotated map of the
      outer halo.}  As in Extended Data Fig. 3 (right panel), now
    shown with the predicted orbit of the LMC (solid white line), a
    line at Dec.$=-25^\circ$ (grey dashed line, marking the
    approximate limit of Northern hemisphere surveys), and the
    locations of the two regions used to measure density ratios in
    Fig. 2 (solid yellow lines).  The yellow region in the North
    measure the collective response while the yellow region in the
    South measures the local wake.}
\end{figure}

\begin{figure*}[t!]
  \includegraphics[width=0.5\textwidth]{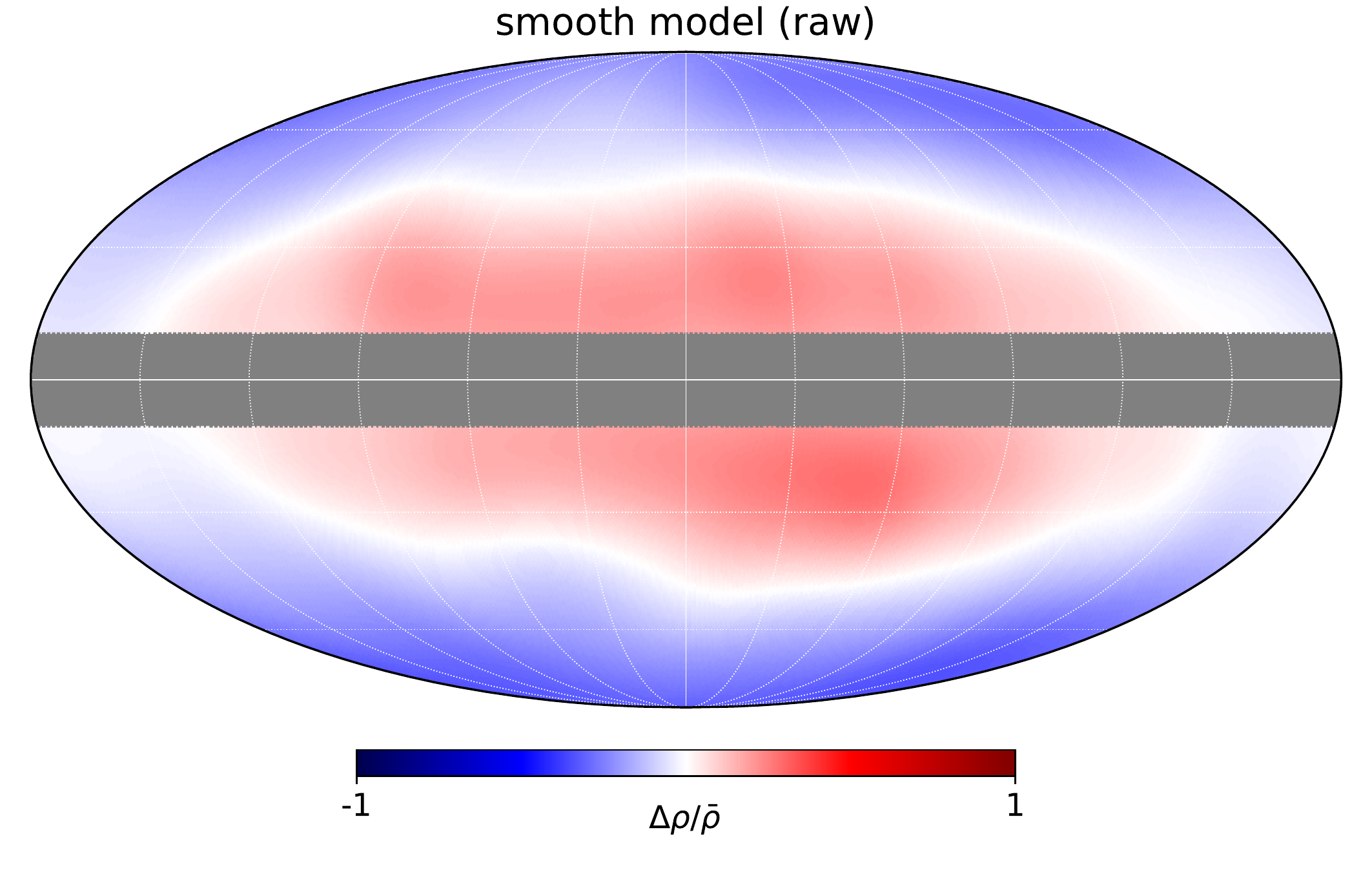}
  \includegraphics[width=0.5\textwidth]{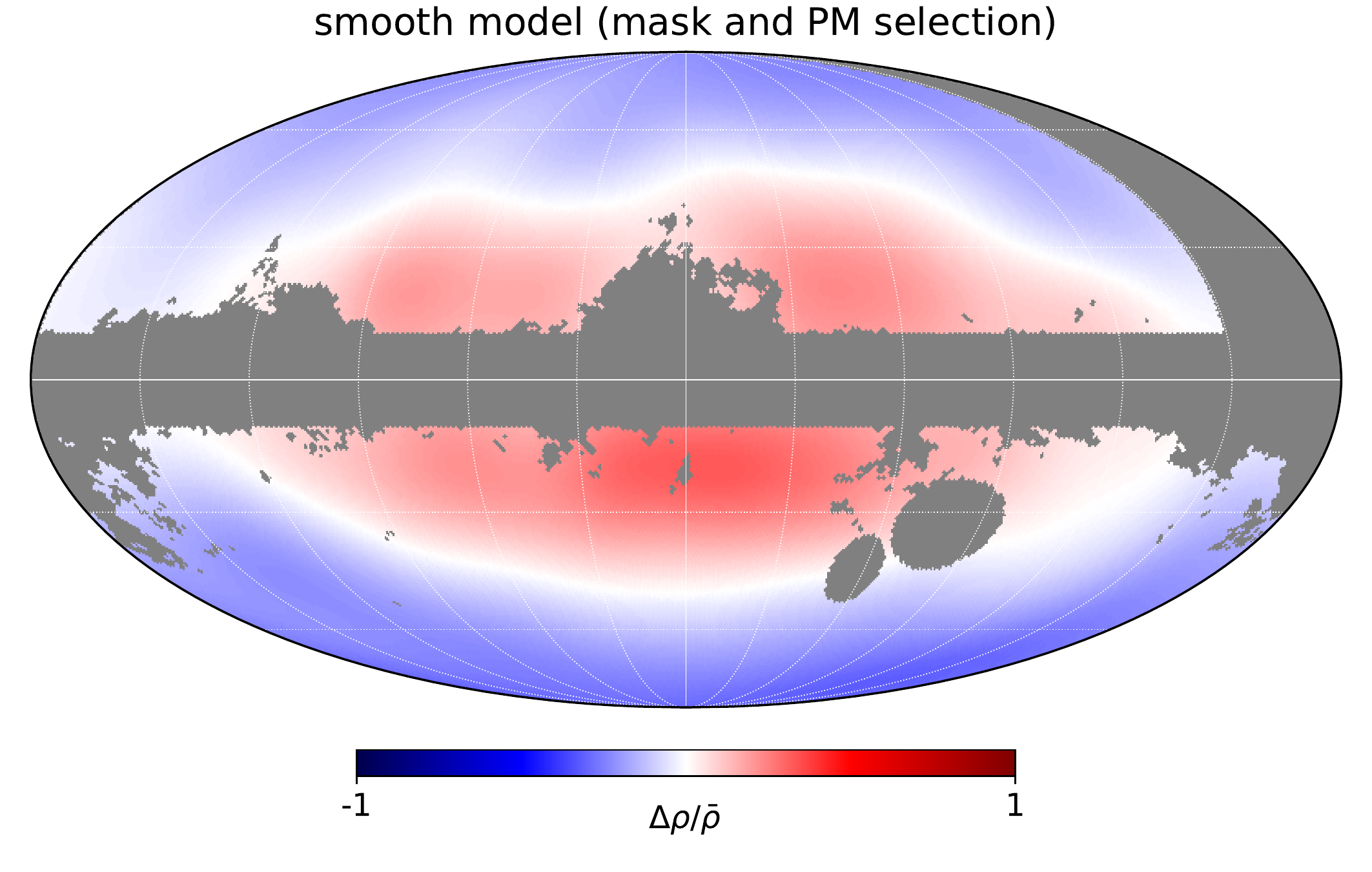}
  \caption{\textbf{Extended Data Figure 6 $|$ Predicted density
      distribution of a smooth model.}  The model\cite{Rybizki18} has
    a smooth (oblate) stellar halo.  {\bf Left panel:} the unfiltered
    smooth model. {\bf Right panel} the smooth model with the same
    selection criteria as used in the data, including various
    coordinate and proper motion cuts.  Both maps have been smoothed
    by a Gaussian with FWHM=$30^\circ$.  Unlike the data this map
    shows no obvious structure. }
\end{figure*}

\vspace{1cm}
\noindent
{\bf The stellar halo probed by RR Lyrae}
\\
\noindent
If the observed overdensities in the Northern and Southern hemispheres
are indeed due to the dynamical effect of the LMC, then a clear
prediction is that the overdensities should be visible in all distant
halo populations that probe the ``smooth'' halo (i.e., neglecting
structures obviously associated with unrelaxed debris).  Here we test
this prediction using RR Lyrae.  These stars are low metallicity
variable stars whose periods can be used to infer distances to a
precision of $\approx3$\%\cite{Sesar17a}.  Here we use the public RR
Lyrae catalog based on Pan-STARRS data\cite{Sesar17a} that reach
distances of $>100$ kpc.

We select stars with $60<R_{\rm gal}<100$ kpc, focusing on the RRab
subtype with a score of $>0.8$, which have more accurate parameters.
The resulting map is shown in Extended Data Fig. 7, which can be
compared directly to Extended Data Fig. 3 (right panel).  There are
overall more RR Lyrae than K giants (as we have defined them here), so
the range of the color bar is larger in the former figure.  The
Pan-STARRS Survey is restricted to Dec.$>-30^\circ$, so the map is
missing data in the lower right quadrant.  We have also masked
$|b|<10^\circ$, regions with E(B-V)$>0.3$, and
$|b_{\rm Sgr}|<15^\circ$ (for $b>0^\circ$).  This last selection was
imposed because the RR Lyrae are too faint to have high quality {\it
  Gaia} proper motions ($G\approx20$), and as a consequence we could
not excise the Sagittarius stream from the maps with the proper motion
cuts used for the K giants.  In this map one sees clearly the local
wake in the Southern hemisphere (first reported as the Pisces
Plume\cite{Belokurov19a}) at a location that is in good
agreement with the K giant sample.  Owing to the Dec$=-30^\circ$ limit
and the Sagittarius mask the RR Lyrae do not probe the overdensity at
$b>0^\circ$, $-180^\circ<l<0^\circ$ as completely as in the K giant
sample.  Nonetheless, an overdensity is clearly visible in that
region.

We provide a quantitative comparison between the RR Lyrae and K giant
maps in the right panel of Extended Data Fig. 7 (cf. Fig. 2).  Here we
have computed the density contrast in the local wake and collective
response regions, including only those pixels not masked.  Because the
RR Lyrae map is much less complete than the K giants, it is not
possible to estimate a robust global mean density from the former.  We
therefore normalize the RR Lyrae map to the K giant map using a large
region that is well-sampled by both maps ($30^\circ<b<60^\circ$,
$30^\circ<l<120^\circ$).  The density contrast of the collective
response computed from the RR Lyrae is statistically indistinguishable
from the measurement based on the K giant sample.  However, the local
wake is 50\% stronger in the RR Lyrae sample.  The reasons for this
are unclear.  The distance distribution of the RR Lyrae is weighted
toward greater distances than the K giants, so the maps are not
probing exactly the same radial distribution.  We caution that the
densities are sensitive to our renormalization procedure.

Further progress could be made employing an all-sky map of RR Lyrae.
{\it Gaia} Data Release 2 provided such a sample\cite{Rimoldini19,
  Clementini19}.  However, these maps are incomplete at the depths
necessary to detect RR Lyrae at $>60$ kpc ($G\approx20$), driven
largely by the {\it Gaia} scanning pattern\cite{Mateu20}.  For
example, the local wake is not clearly visible in the {\it Gaia} RR
Lyrae catalog despite the fact that it is so prominent in the
Pan-STARRS data at $>60$ kpc.  Future data releases from {\it Gaia}
will hopefully reach the depths necessary to deliver complete samples
of RR Lyrae in the outer halo.

\vspace{.3cm}
\noindent
{\bf Simulation details}
\\
\noindent
The simulations shown here were initialized with a spherical dark
matter halo following a Hernquist profile\cite{Hernquist90}, a stellar
disk, and a stellar bulge with masses of
$1.54\times 10^{12}\,M_{\odot}$, $5.78\times10^{10}\,M_{\odot}$, and
$0.9\times10^{10}\,M_{\odot}$ respectively. Each particle has a mass
of $m_p=1.57\times10^4\,M_{\odot}$. One model of the MW has isotropic
halo kinematics while the other has a radially biased kinematic
profile. The 4 LMC models were initialized with a spherical dark
matter Hernquist profile with total halo masses of 0.8, 1, 1.8,
$2.5\times10^{11} \,M_{\odot}$. The scale length of each of the LMC
halos was chosen to match the observed rotation curve of the LMC
within 9 kpc. Detailed parameters for these simulations can be found
in Table 1 of the original simulation paper\cite{Garavito-Camargo19}.
All the initial conditions for the halos were built using
GalIC\cite{Yurin14}. The N-body simulations were run with P-gadget3.
The initial conditions for the orbit of the LMC were found iteratively
until the present-day position and velocity of the LMC was within
$2\sigma$ of the present-day observed properties of the LMC.

In this work we have utilized the raw dark matter particle data to
apply a simple re-weighting scheme to match the observed density
profile.  The left panel of Extended Data Fig. 8 shows the $N-$body
simulation presented in Fig. 1, now without any selections, aside from
matching the density profile of the data and selecting stars with
$60<R_{\rm gal}<100$ kpc.    The right panel shows the same simulation
now viewed from an observer placed at the Galactic center.  This
perspective provides a sharper distinction between the local wake and
collective response\cite{Garavito-Camargo19}, but we opted not to make
the observed maps in this way owing to the complex mapping of the Galactic
reddening selection in this projection.

Another approach to constructing a mock stellar halo from these
simulations is available for these
simulations\cite{Garavito-Camargo19} and was built in equilibrium with
the dark matter halo, given a specified stellar density and velocity
dispersion profile using a weighing scheme designed to reproduce the
observed density profile of the stellar halo.  We have compared our
mock stellar halo to this alternative approach and find very similar
results.

The detailed structure of the halo response to the passage of the LMC
has been recently quantified using Basis Function
Expansions\cite{Garavito-Camargo20}. The halo response to the LMC has
a strong amplitude in odd $l$ modes. In contrast, triaxial, oblate and
prolate halos have no odd $l$ terms, only even terms.  As such, the
deformations to the MW halo caused by the LMC cannot be mimicked by a
triaxial halo.

\vspace{1cm}
\noindent
{\bf A tilted halo?}
\\
\noindent
The observed all-sky map of halo stars shown in Fig. 1a displays a
degree of symmetry in the sense that the overdensities are confined to
the south-west and north-east.   We explore in this section the
possibility that the observed density variation could be due to a
smooth tilted stellar halo instead of our preferred interpretation of
the impact of the LMC.

Extended Data Fig. 9 shows a smooth, tilted triaxial halo whose
parameters were optimized (by hand) to mimic the observations in
Fig. 1a.  Specifically, the model has axial ratios of 0.2
(minor-to-major axes) and 0.6 (intermediate-to-major axes) and is
rotated $60^\circ$ anti-clockwise along the y-axis.  Note that while
this model reproduces some of the basic features of the data, in
detail there are significant differences in both the Southern and
Northern hemispheres that leads us to disfavor this interpretation of
the data.  Moreover, independent results based on kinematics of the
stellar halo also favor the LMC dynamical response interpretation of
the observed outer stellar halo\cite{Petersen20b, Erkal20b}.

\begin{figure*}[t!]
  \includegraphics[width=0.5\textwidth]{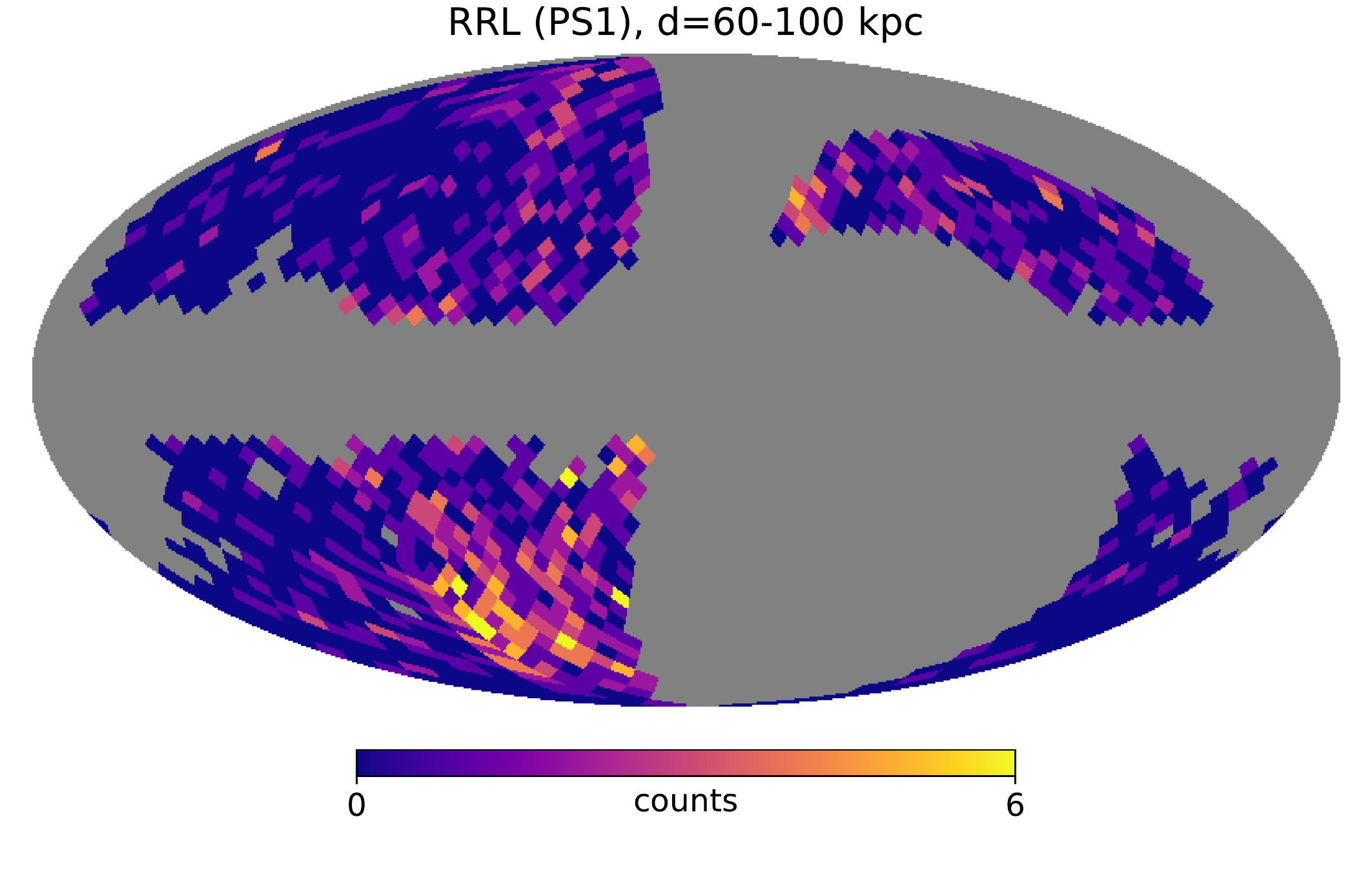}
  \includegraphics[width=0.5\textwidth]{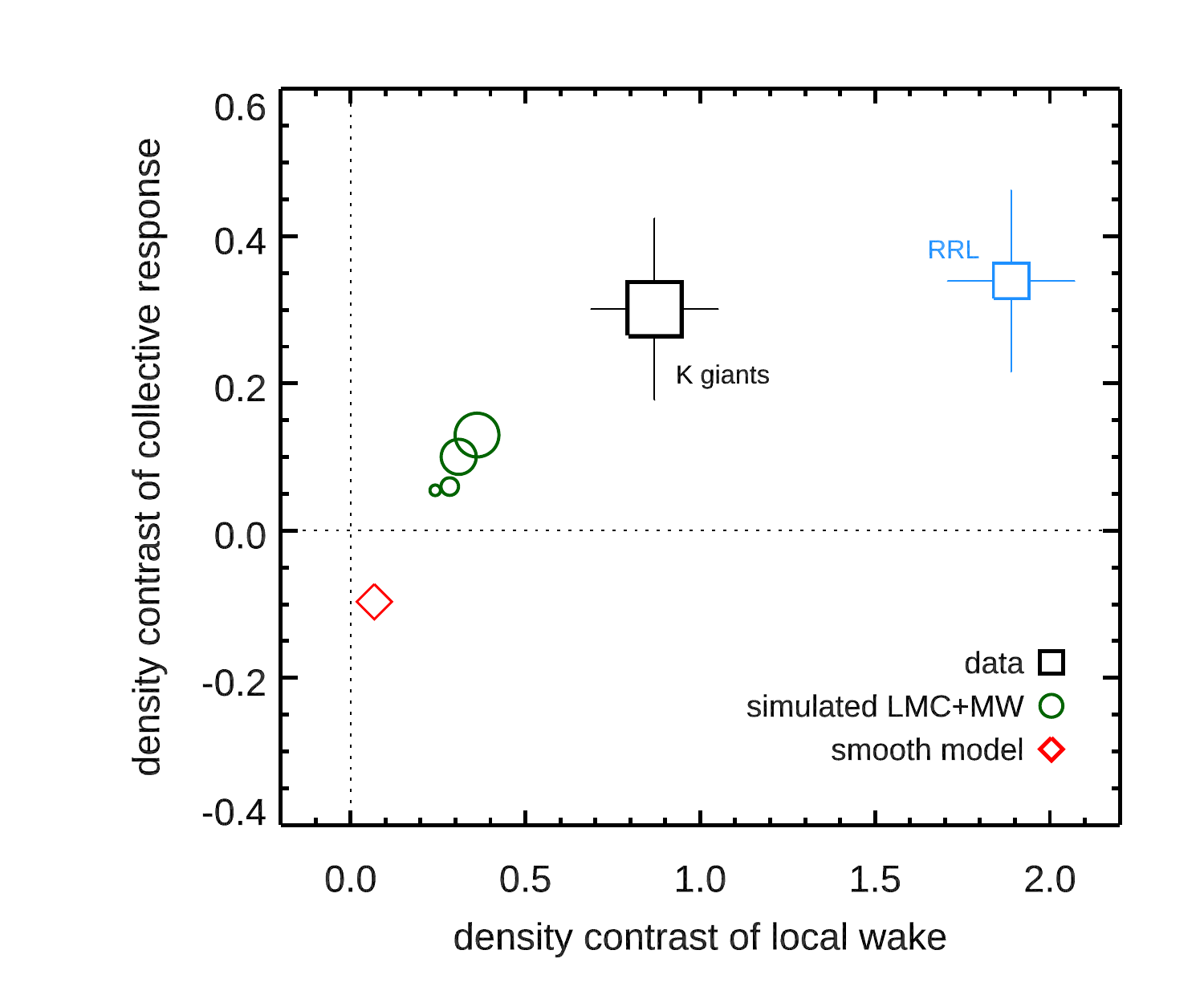}
  \caption{\textbf{Extended Data Figure 7 $|$ RR Lyrae as a probe of
      the stellar halo.}  {\bf Left panel:} the binned all-sky map of
    RR Lyrae stars identified in Pan-STARRS data \cite{Sesar17a}.  The
    data are restricted to Dec.$>-30^\circ$.  The wake is clearly
    visible in the lower left quadrant (compare with Extended Data
    Fig. 4).  {\bf Right panel:} measured densities in the wake and
    collective response regions for RR Lyrae (blue) compared to the K
    giants (black; compare with Fig. 2).  Both the wake and collective
    response are clearly detected in the RR Lyrae.  }
\end{figure*}

\begin{figure*}[t!]
  \includegraphics[width=0.5\textwidth]{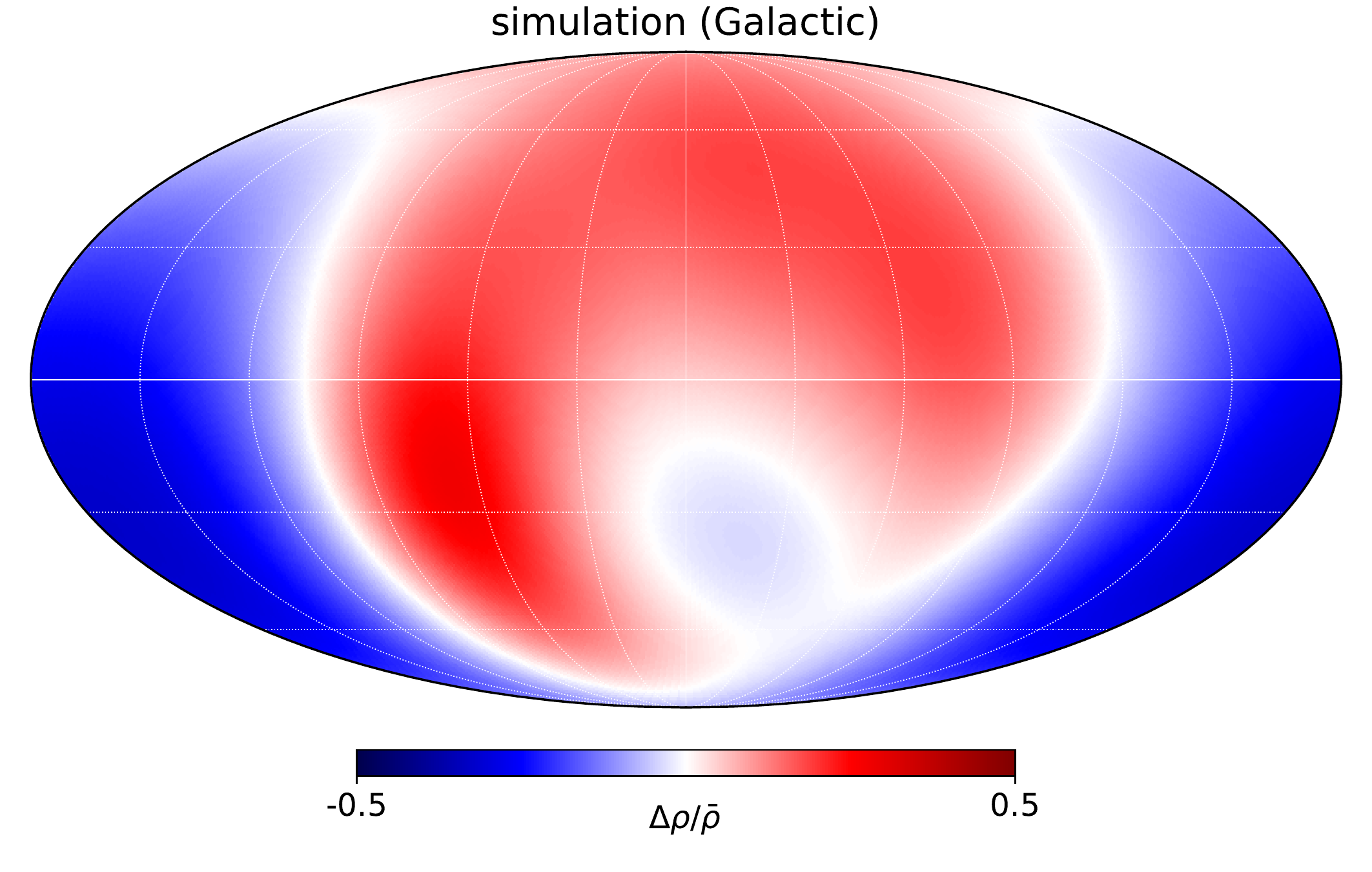}
  \includegraphics[width=0.5\textwidth]{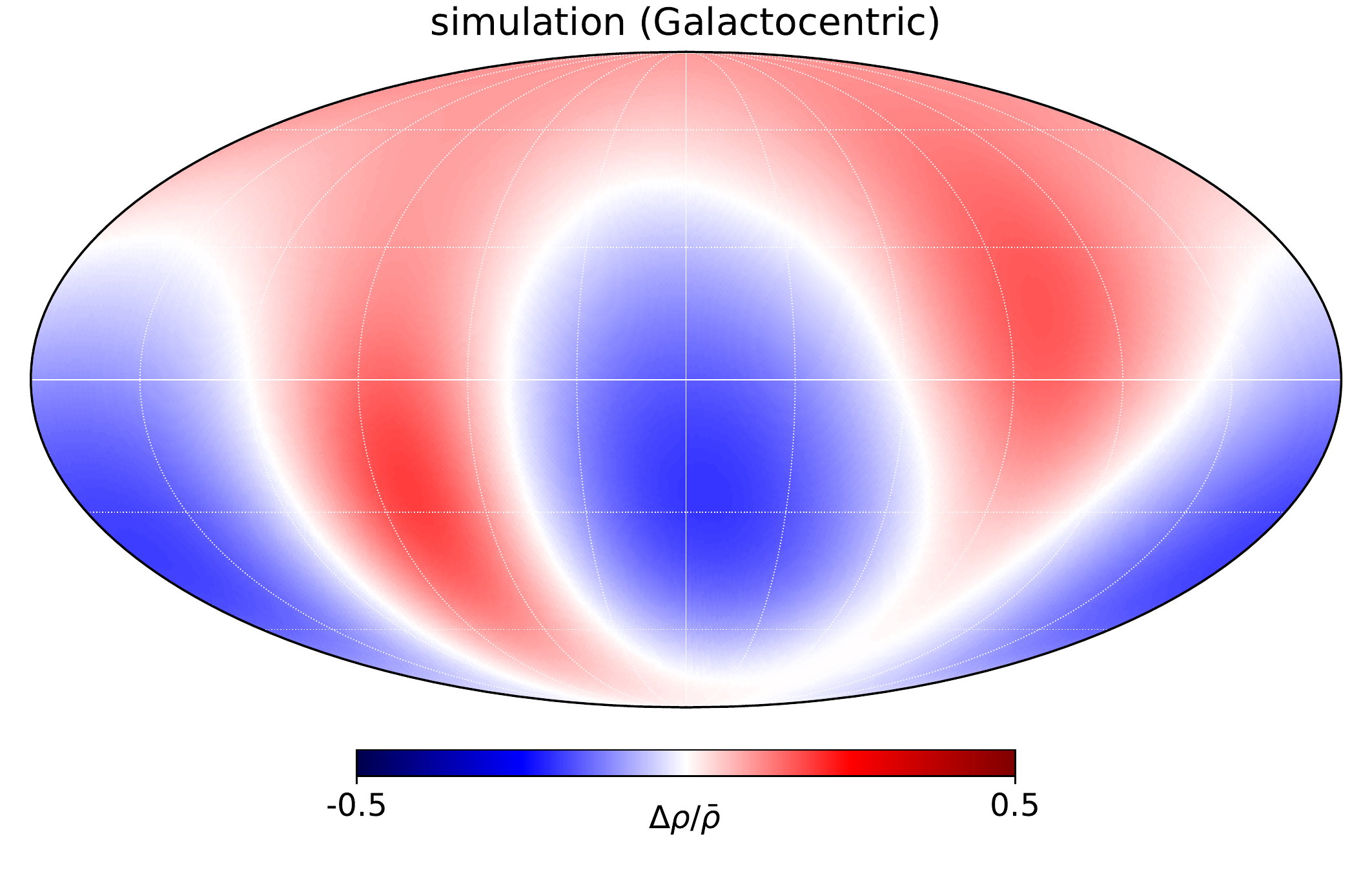}
  \caption{\textbf{Extended Data Figure 8 $|$ Predicted response of
      the Galactic halo to the LMC.}  The $N-$body simulation
    presented in Fig. 1b is shown here without any selections, either
    in proper motions or on-sky (the density profile is still matched
    to the data, and only stars with $60<R_{\rm gal}<100$ kpc are
    included). {\bf Left panel:} projection in the usual Galactic
    coordinates.  {\bf Right panel:} projection in Galactocentric
    coordinates (what an observer would see if placed at the Galactic
    center).  In the right panel the model amplitude of the collective
    response is asymmetric and is largest near the Galactic
    plane.}
\end{figure*}

\begin{figure}[t!]
  \includegraphics[width=0.5\textwidth]{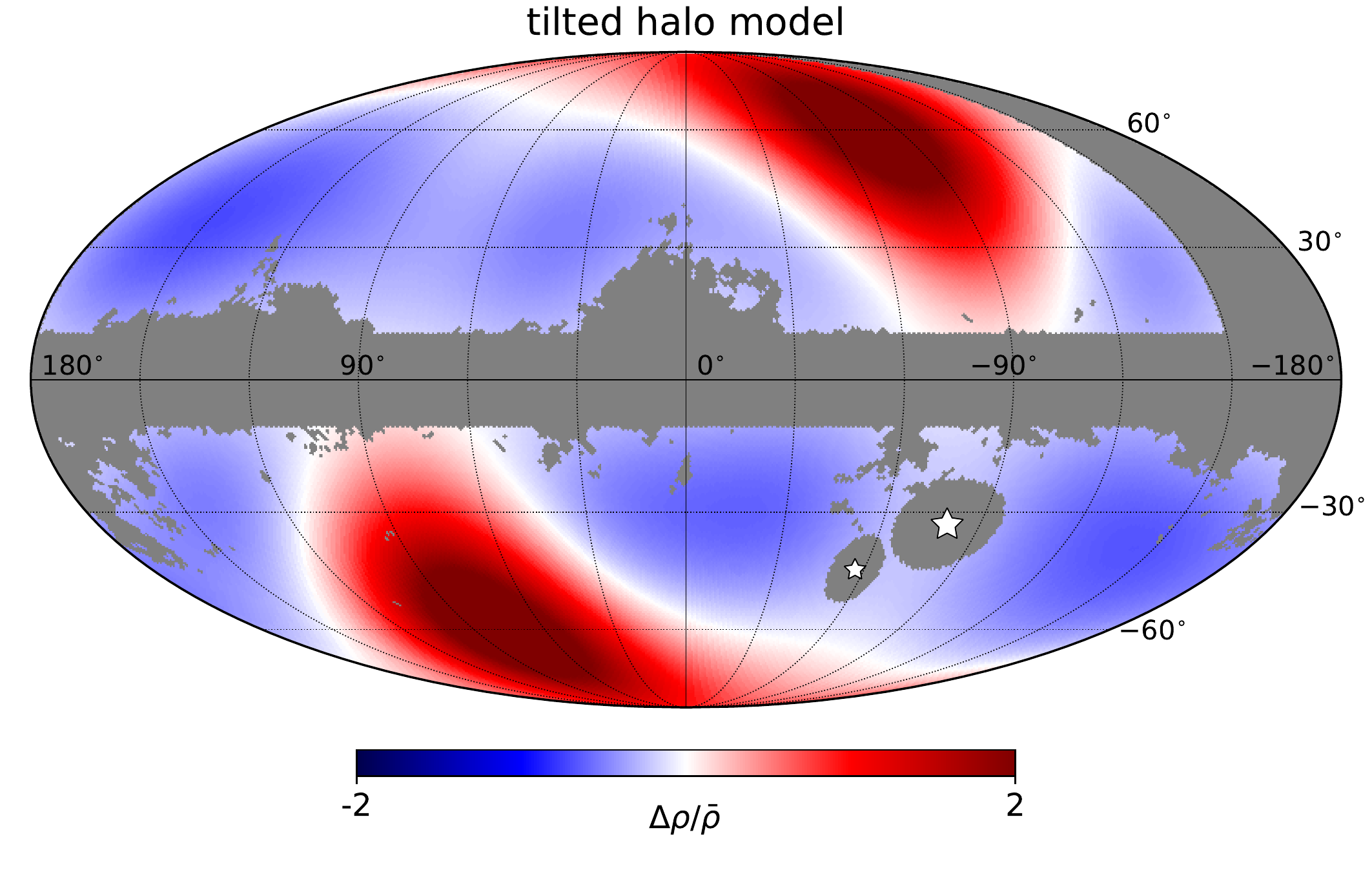}
  \caption{\textbf{Extended Data Figure 9 $|$ Predicted density
      distribution of a tilted stellar halo.}  All-sky density
    distribution of a smooth triaxial model stellar halo that is
    tilted by $60^\circ$ along the y-axis.  While this model captures
    some of the qualitative behavior seen in the data (Fig 1a), it
    fails to reproduce both the detailed shape of the local wake and
    predicts a precise symmetry in the North and South, which is not
    observed. }
\end{figure}

\end{document}